\documentclass[
  aps,
  prl,
  twocolumn,
  superscriptaddress,
  amsmath,
  amssymb,
  floatfix,
  10pt
]{revtex4-2}

\usepackage{graphicx}
\usepackage{bm}
\usepackage{physics}
\usepackage{titlesec}
\usepackage[version=4]{mhchem}
\usepackage{hyperref}
\hypersetup{colorlinks=true, citecolor=blue, urlcolor=blue, linkcolor=blue}

\usepackage{xcolor}

\begin{document}

\title{Dynamic Screening Effects on Auger Recombination in Metal-Halide Perovskites}

\author{Utkarsh Singh\footnotemark}
\thanks{Corresponding author: \href{mailto: utkarsh.singh[at]liu.se}{utkarsh.singh[at]liu.se}}
\affiliation{%
  Theoretical Physics Division, Department of Physics, Chemistry, and Biology (IFM),\\
  Link\"opings Universitet, SE-581 83 Link\"oping, Sweden
}

\author{Sergei I. Simak}
\affiliation{%
  Theoretical Physics Division, Department of Physics, Chemistry, and Biology (IFM),\\
  Link\"opings Universitet, SE-581 83 Link\"oping, Sweden
}
\affiliation{%
  Department of Physics and Astronomy, Uppsala University, SE-75120 Uppsala, Sweden
}

\date{\today}

\begin{abstract} 
    The performance of modern light-emitting technologies, from lasers to LEDs, is limited by nonradiative losses, with Auger recombination being the dominant channel at device-relevant carrier densities. Reliable modeling of this process is essential, yet conventional treatments neglect dynamic dielectric effects, limiting the predictive reliability at operating conditions. We develop a general framework that incorporates the frequency-dependent screened Coulomb interaction $W_{00}(\mathbf{q},\omega)$, computed from low-scaling \textit{GW}, into both direct and phonon-assisted Auger amplitudes. Demonstrated on orthorhombic $\gamma$-CsPbI$_3$ (band gap $E_g\approx1.73$ eV) and $\gamma$-CsSnI$_3$ ($E_g\approx1.30$ eV), the approach shows that dynamic screening enhances the dielectric response, lowering the room-temperature Auger coefficient by $\sim$50-60 \%. This renormalization shifts the crossover between radiative and nonradiative recombination by nearly a factor of two in carrier density. Dynamic dielectric screening thus emerges as a quantitative determinant of Auger recombination, offering a transferable framework for predictive modeling across polar semiconductors where 
    frequency-independent screening models are inadequate.
\end{abstract}
    
\maketitle
Understanding and predicting non-radiative recombination processes is essential for advancing the performance of next-generation optoelectronic devices, particularly under high excitation conditions where such losses become increasingly detrimental. Among these processes, Auger recombination (AR) scales with the third power of the carrier density,
\(R_{\mathrm{AR}}\propto n^3\), and therefore becomes the dominant non-radiative
loss at high injection conditions commonly encountered in perovskite nanolasers and in high-brightness light-emitting diodes operated at current densities of order of kA\,cm\(^{-2}\) .%
\cite{Shen2018AEM,Qin2021TrChem,Sun2024APR,Zou2020ACSNano}

Time-resolved photoluminescence and pump-probe studies on hybrid and all-inorganic lead-halide perovskites consistently extract Auger coefficients
\(C \sim 10^{-29}\text{-}10^{-28}\,\mathrm{cm^{6}\,s^{-1}}\),%
\cite{PazosOuton2018JPCL,Bowman2021ACSEnergy}
one to two orders larger than the \(10^{-30}\text{-}10^{-31}\,\mathrm{cm^{6}\,s^{-1}}\) characteristic of typical device-use semiconductors in GaAs and Si.%
\cite{Govoni2011PRB,Steiauf2014ACSPho}
First-principles calculations reproduced the large Auger coefficient \(C\) reported for MAPbI\(_3\), attributing it to a near-resonance between the band gap and higher conduction states that is further enhanced by spin-orbit coupling,%
\cite{Shen2018AEM,Zhang2020AEM}.

However, the Coulomb kernel in those studies employed model dielectric functions  \(\varepsilon(\mathbf{q})\) parametrized by the ion-clamped constant \(\varepsilon_\infty\), with no frequency dependence.\cite{Kioupakis2015PRB}
This approximation neglects the frequency dependence, \(\varepsilon(\omega)\), that can be critical for ultrafast electronic processes such as Auger recombination.

In polar iodide perovskites, the ionic lattice cannot respond on femtosecond time scales, and the interband (electronic) polarization varies appreciably across the relevant range (0-5 eV), so \(\varepsilon^{-1}(\mathbf{q},\omega)\) can differ markedly from its 
ion-clamped and static 
limits in the spectral window relevant to Auger final-state phase space.%
\cite{Herz2018JPCL,Leveillee2019PRB}
Dynamic screening is already known to substantially reshape carrier-carrier scattering and Auger(-Meitner) rates in other materials -- order-of-magnitude effects in graphene when many-body screening is 
incorporated, and strong rate changes in conventional semiconductors.%
\cite{Alymov2018PRB}
Yet its impact on Auger recombination in soft, metal-halide perovskites remains largely unexplored.

\begin{figure}[htbp]
    \centering
    \includegraphics[width=\columnwidth]{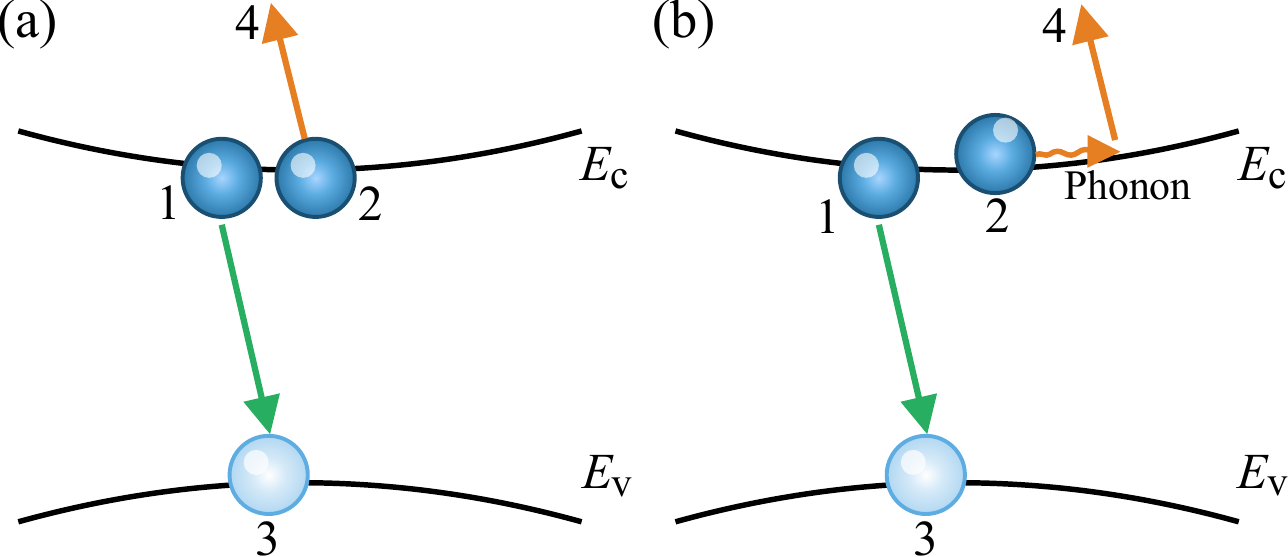}
    \caption{Schematic illustration of 
    (a) Direct Auger recombination, and
    (b) Indirect (phonon-assisted) Auger recombination 
    \textit{E}$_\mathrm{c}$ and \textit{E}$_\mathrm{v}$ mark the conduction band minima and valence band maxima, respectively. 
    The opposing vectors in both processes signify conservation of energy and crystal momentum for the overall transition. The state indices 1, 2, 3 and 4 dictate the convention used throughout the text.}
    \label{fig0}
\end{figure}

We address this fundamental discrepancy by introducing a general first-principles framework that incorporates the fully \emph{frequency-dependent} screened Coulomb interaction into Auger recombination calculations. To demonstrate its impact, we apply the framework to the orthorhombic \(\gamma\) phases of CsPbI\(_3\) (\(E_g\!\approx\!1.73~\mathrm{eV}\)) and CsSnI\(_3\) (\(\approx\!1.30~\mathrm{eV}\)). 
Although the two compounds have the same corner-sharing \(BX_6\) framework, they differ in band gap, spin-orbit coupling strength, and dielectric dispersion, providing a clean \(B\)-site (Pb \(\rightarrow\) Sn) test bed for dynamic-screening effects. Although the two compounds have the same corner-sharing \(BX_6\) framework, they differ in band gap, spin-orbit coupling strength, and dielectric dispersion, providing a clean \(B\)-site (Pb \(\rightarrow\) Sn) test bed for dynamic-screening effects.

Accounting for the macroscopic head of the screened interaction, $W_{00}(\mathbf{q},\omega)$ (i.e., $G=G'=0$), suppresses long-range Coulomb scattering at optical energies and \emph{lowers} the room-temperature Auger coefficient by $50$-$60$\,\% relative to the \emph{frequency-independent} baseline $W_{00}(\mathbf q,0)$ (whose long-wavelength head is $v(\mathbf q)/\varepsilon_\infty$).

Taken together, these benchmarks could narrow the gap to experimental observations and point to dielectric- and phonon-engineering routes for mitigating Auger losses in perovskite emitters.

\paragraph{Electronic structure}
For both materials under study, electronic-structure calculations at the 
$G_{0}W_{0}$ level were performed using the low-scaling $GW$ algorithm, yielding band gaps in agreement with experiment 
to within 0.10\,eV. A set of 64 spinor, maximally localized Wannier functions, spanning [\textsc{\small VBM$-6$, CBM$+6$}] \,eV was constructed, reproducing the $G_{0}W_{0}$ 
band structure with an accuracy better than 10\,meV. Diagonalizing the 
real-space Hamiltonians 
then provided $G_{0}W_{0}$ 
eigenvalues on dense $60 \times 40 \times 60$ $k$-mesh, along with Bloch 
overlaps, which serve as inputs for the Auger phase-space integrations as detailed in the Supplementary Information (S.I.).

\begin{figure*}[t]
  \centering
  \includegraphics[width=\linewidth]{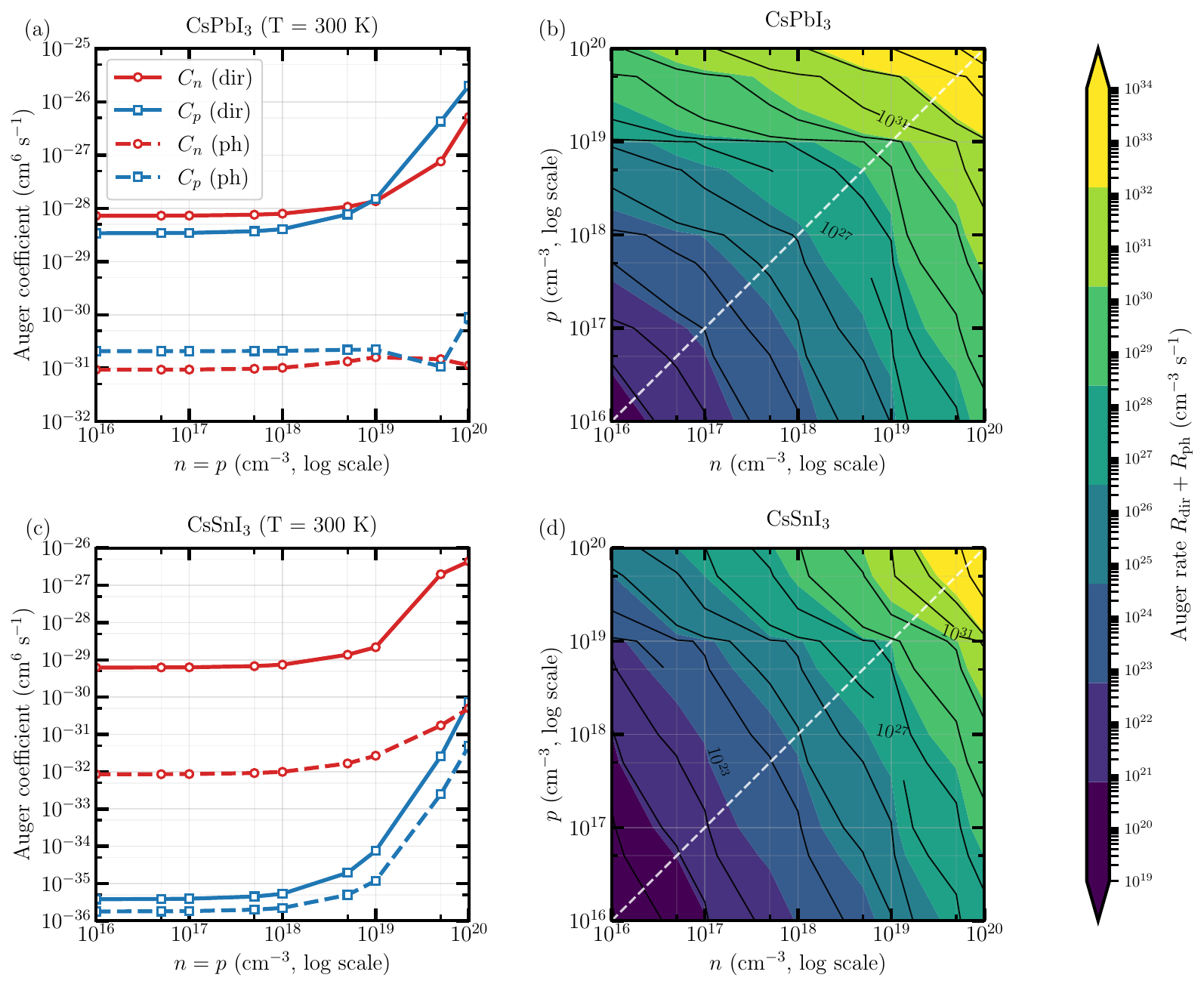}
    \caption{%
    (a,c) Auger coefficients along the $n=p$ locus at $T=300\,\mathrm{K}$ for \ce{CsPbI3} (a) and \ce{CsSnI3} (c).
    Solid lines show direct coefficients (red: $C_n$ (\text{eeh}), blue: $C_p$ (\text{hhe})). Dashed lines show the phonon-assisted counterparts.
    (b,d) Dynamically screened total Auger recombination rate $R_{\mathrm{tot}}=R_{\mathrm{dir}}+R_{\mathrm{ph}}$ as a function of electron ($n$) and hole ($p$) densities for \ce{CsPbI3} (b) and \ce{CsSnI3} (d).
    All axes are logarithmic.
    }
  \label{fig1}
\end{figure*}

\paragraph{Frequency-dependent screening.}
We evaluate the dynamically screened head
\begin{equation}
\begin{aligned}
  W_{00}(\mathbf{q},\omega)
  = \sum_{G'} \varepsilon^{-1}_{0G'}(\mathbf{q},\omega)\,v(\mathbf{q}+\mathbf{G}').
\end{aligned}
\label{eq:W-main}
\end{equation}
where   $\quad v(\mathbf{q})=\frac{4\pi}{|\mathbf{q}|^{2}}$. 
The frequency argument $\omega$ corresponds to the electronic energy transfer involved in a given Auger scattering event, for example, the energy difference between initial and final electronic states in the direct and exchange channels (detailed expressions in the S.I.).
We compute $\varepsilon^{-1}_{00}(\mathbf q,i\xi)$ on the imaginary axis within the low-scaling $G_0W_0$ framework and analytically continue it to $\omega{+}i\eta$ using a passive (non-negative) pole expansion fitted to the Matsubara data.

At long wavelength we use the macroscopic limit $W_{00}\simeq v(\mathbf q)/\varepsilon_M(\omega)$ (where $\varepsilon_M$ is the macroscopic dielectric function), and away from this limit we evaluate the microscopic head as $W_{00}(\mathbf q,\omega)=\varepsilon^{-1}_{00}(\mathbf q,\omega)\,v(\mathbf q)$.
For comparison, the \emph{frequency-independent baseline} uses 
$W_{00}(\mathbf{q})\equiv W_{00}(\mathbf{q},\omega\!=\!0)=\varepsilon^{-1}_{00}(\mathbf{q},0)\,v(\mathbf{q})$, which in the macroscopic limit ($|\mathbf{q}|\!\to\!0$) reduces to $v(\mathbf{q})/\varepsilon_\infty$, where $\varepsilon_\infty$ denotes the ion-clamped (electronic) \emph{macroscopic} dielectric constant.
While the baseline reduces to $\varepsilon_\infty$ at long wavelength, it includes full $q$-dependent local-field effects at finite momentum.
Throughout the text we refer to this as the \emph{baseline} calculation.
Details of sampling, continuation, $\Gamma$-cell averaging, and passivity/Kramers-Kronig checks are provided in Sec. S11 of the S.I..

\paragraph{Auger matrix elements and rate.}
The evaluation of Auger recombination rates, involving electron-initiated ($eeh$), hole-initiated ($hhe$), and phonon-assisted channels, has been discussed extensively in the previous works \cite{Bushick2023PRL, Kioupakis2015PRB}. In the present work, we provide a discretized formulation tailored to our calculations, highlighting that all relevant quantities are represented on a finite $n$-dimensional grid. 
The following framework allows us to compute the direct Auger recombination rate for the direct $eeh$ mechanism. Using Fermi’s golden rule,
\begin{equation}
\begin{aligned}
R_{eeh}^{\mathrm{dir}}
&=\frac{2\pi}{\hbar}\,\frac{V_{\mathrm{BZ}}^{3}}{\Omega}\,
\Big\langle \delta_{\sigma}\!\bigl(\Delta E\bigr)\;
P_{eeh}\;\bigl|M_{eeh}\bigr|^{2}\Big\rangle_{\mathcal{Q}},\\
\Delta E&=\epsilon_{1}+\epsilon_{2}-\epsilon_{3}-\epsilon_{4},\\
P_{eeh}&=f_c(\epsilon_{1})\,f_c(\epsilon_{2})\,
  \bigl[1-f_v(\epsilon_{3})\bigr]\,
  \bigl[1-f_c(\epsilon_{4})\bigr].
\end{aligned}
\label{eq:Rdir-main}
\end{equation}
where $\langle\cdots\rangle_{\mathcal{Q}}$ denotes a weighted average over momentum- and energy-conserving
quadruplets $\mathcal{Q}$, sampled on dense $k$ meshes (explicit Monte Carlo estimators in Sec. S9 of the S.I.), and $\delta_{\sigma}$ is a normalized energy kernel. $V_{\mathrm{BZ}}$ is the Brillouin zone volume and $\Omega$ is the unit cell volume. The state indices follow the standard Auger convention illustrated in Fig.~\ref{fig0}.
The antisymmetrized matrix element is
\begin{equation}
\begin{aligned}
M_{eeh}&=M_{eeh}^{\mathrm{dir}}-M_{eeh}^{\mathrm{exc}},\\
M_{eeh}^{\mathrm{dir}}
&=W_{00}\!\bigl(\mathbf{q}_{41},\omega_{41}\bigr)\;
\underbrace{\braket{u_{4\mathbf{k}_4}}{u_{1\mathbf{k}_1}}}_{O_{41}}\,
\underbrace{\braket{u_{2\mathbf{k}_2}}{u_{3\mathbf{k}_3}}}_{O_{23}},\\
M_{eeh}^{\mathrm{exc}}
&=W_{00}\!\bigl(\mathbf{q}_{31},\omega_{31}\bigr)\;
\underbrace{\braket{u_{3\mathbf{k}_3}}{u_{1\mathbf{k}_1}}}_{O_{31}}\,
\underbrace{\braket{u_{2\mathbf{k}_2}}{u_{4\mathbf{k}_4}}}_{O_{24}},
\end{aligned}
\label{eq:Meeh-main}
\end{equation}
where $u_{n\mathbf{k}}$ is the periodic part of the Bloch state, with overlaps $O_{ij}$ evaluated from Wannier-interpolated wavefunctions (detailed in Sec. 10 of S.I.). The transferred momenta are $\mathbf{q}_{41}=\mathbf{k}_4-\mathbf{k}_1+\mathbf{G}_{41}$ and
$\mathbf{q}_{31}=\mathbf{k}_3-\mathbf{k}_1+\mathbf{G}_{31}$ (Umklapp allowed), and bosonic
frequencies set by electronic energy differences,
$\hbar\omega_{41}=\epsilon_{4}-\epsilon_{1}$ and
$\hbar\omega_{31}=\epsilon_{3}-\epsilon_{1}$. The $hhe$ mechanism channel is obtained by the
interchange $c\!\leftrightarrow\!v$ with analogous definitions.
The structure for phonon-assisted Auger rates $R_{eeh}^{\mathrm{ph}}$, 
is analogous to Eq.~(\ref{eq:Rdir-main}) with
$\delta_{\sigma}\!\bigl(\Delta E\bigr)\!\to\!\delta_{\sigma}\!\bigl(\Delta E-s\hbar\Omega_{\mathbf{q}\nu}\bigr)$
and e-ph matrix elements entering $|M|^{2}$; full expressions and sampling details are laid out in the S.I.

The conventional coefficients reported in the main text are
\begin{equation}
\begin{aligned}
C_n(T;n,p)=\frac{R_{eeh}(T;n,p)}{n^2 p},\\
C_p(T;n,p)=\frac{R_{hhe}(T;n,p)}{p^2 n}
\end{aligned}
\label{eq:CnCp-main}
\end{equation}
with $R_{eeh}=R_{eeh}^{\mathrm{dir}}{+}R_{eeh}^{\mathrm{ph}}$ and
$R_{hhe}=R_{hhe}^{\mathrm{dir}}{+}R_{hhe}^{\mathrm{ph}}$.

Figure \ref{fig1} summarizes the dynamically screened Auger landscape in \ce{CsPbI3} and \ce{CsSnI3}. 
It is immediately apparent that at comparable carrier densities, the Auger recombination rate, and thus the coefficient for \ce{CsPbI3} is an order of magnitude or more larger in comparison to \ce{CsSnI3}.


\begin{figure}[t]
\centering
\includegraphics[width=\linewidth]{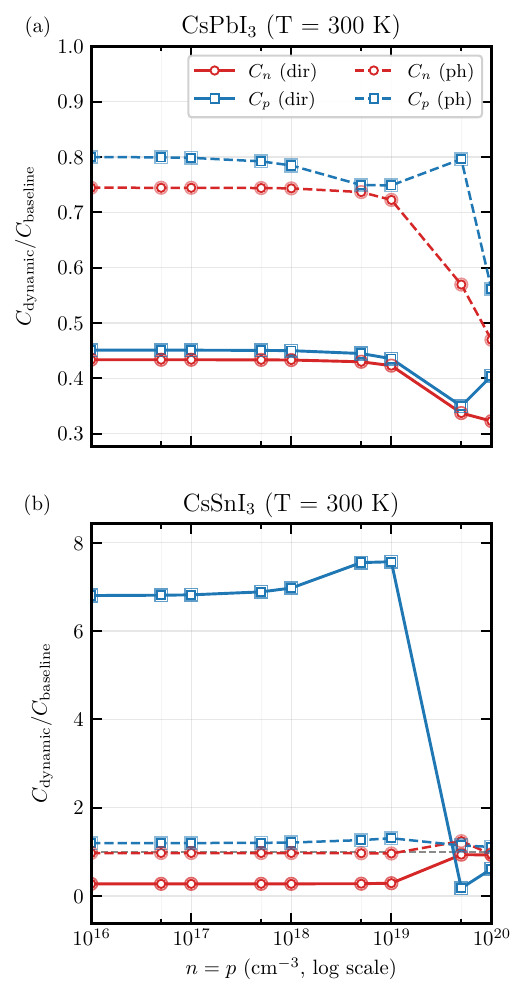}
\caption{%
Ratio of Auger rate coefficients calculated using frequency-dependent (dynamic) and frequency-independent (baseline) screening at
$T=300$~K. (a) CsPbI$_3$, (b) CsSnI$_3$. $C$ (dir) and $C$ (ph) represent the direct and phonon assisted coefficients respectively.
}
\label{fig2}
\end{figure}

This is expected, owing to a smaller phase space volume \cite{Yuan2023} available to \ce{CsSnI3} for Auger events. We observe that for \ce{CsPbI3}, $C_n$ leads $C_p$ by a small margin under device-relevant carrier densities, and there is a reversal of this trend when the $n=p$ locus goes into the degenerate carrier regime of $(n,p)\!\ge\!10^{19}\,\mathrm{cm^{-3}}$. The phonon-assisted Auger coefficients are two to three orders of magnitude lower, which is typical for direct bandgap semiconductors. For \ce{CsSnI3}, in contrast, there is a large deficit in the $C_p$ coefficient with respect to its counterpart, both through direct and phonon-assisted channels, revealing a much smaller probability of Auger events via the $hhe$ mechanism when rate calculations are concerned.

Figures \ref{fig1} (b) and (d) illustrate a more general result, serving as a reference, as it figures in $n \neq p$ conditions, which is the case for most devices in operation. The total Auger rate $R_\mathrm{tot}$ for \ce{CsPbI3} is observed to be higher than that of \ce{CsSnI3} at nearly all reasonable $(n,p)$ points.

\paragraph{Dynamic versus 
baseline screening.}

Figure~\ref{fig2} quantifies the impact of frequency-dependent screening by comparing dynamic and baseline Auger coefficients along the $n=p$ locus.

The additional interband polarization encoded in $\varepsilon(\mathbf{q},\omega)$ increases the effective dielectric response at optical energies and therefore reduces the screened interaction $|W_{00}|$, lowering the coefficients under otherwise identical conditions. This behavior is consistent with many-body theory of dielectric screening and $GW$ response functions,\cite{Onida2002RMP} and with the strongly dispersive electronic dielectric function reported for halide perovskites,\cite{Herz2018JPCL,Wilson2019APLM} as well as prior first-principles demonstrations that frequency-dependent screening can reshape carrier-carrier scattering and Auger-type rates in semiconductors.\cite{Alymov2018PRB}. This lowers the dominant, direct Auger coefficient by $\sim50 - 60$\,\% in $\gamma$-CsPbI$_3$ and $\gamma$-CsSnI$_3$, while the indirect coefficient is affected weakly. The larger drop for the Pb based compound can likely be attributed to its slightly stronger dielectric dispersion.

A notable exception is the hole-initiated coefficient in \ce{CsSnI3}: under dynamic screening $C_p$ \emph{increases}, whereas $C_n$ (and both $C_n$, $C_p$ in \ce{CsPbI3}) decrease. We ascribe this to two concurrent effects. First, because $\gamma$-\ce{CsSnI3} has a smaller gap ($E_g\!\approx\!1.3$\,eV) than $\gamma$-\ce{CsPbI3} ($E_g\!\approx\!1.7$\,eV), the $hhe$ matrix element samples frequencies closer to interband structure where $\varepsilon_M(\omega)$ is strongly dispersive and $\varepsilon_M^{-1}(\omega)$ acquires a significant \emph{complex} part; the resulting magnitude–phase change in $W_{00}(\mathbf q,\omega)$ reweights direct vs.\ exchange and reduces their cancellation in the $hhe$ amplitude. Second, the much weaker spin–orbit coupling in Sn-based perovskites preserves near-degeneracies and larger valence-edge overlaps, making the $hhe$ channel more sensitive to this reweighting, relative to its Pb-counterpart. We emphasize that across the relevant $(n,p)$, the electron-initiated channel remains dominant ($C_n\!\gg\!C_p$) in \ce{CsSnI3}, so the qualitative hierarchy of Auger pathways is unchanged.
No time-resolved Auger data exist for these phases to date, and the numbers
in Fig.~\ref{fig1} and Fig.~\ref{fig2} thus constitute first-principles benchmarks for
device modeling.


\begin{figure}[t]
\centering
\includegraphics[width=\linewidth]{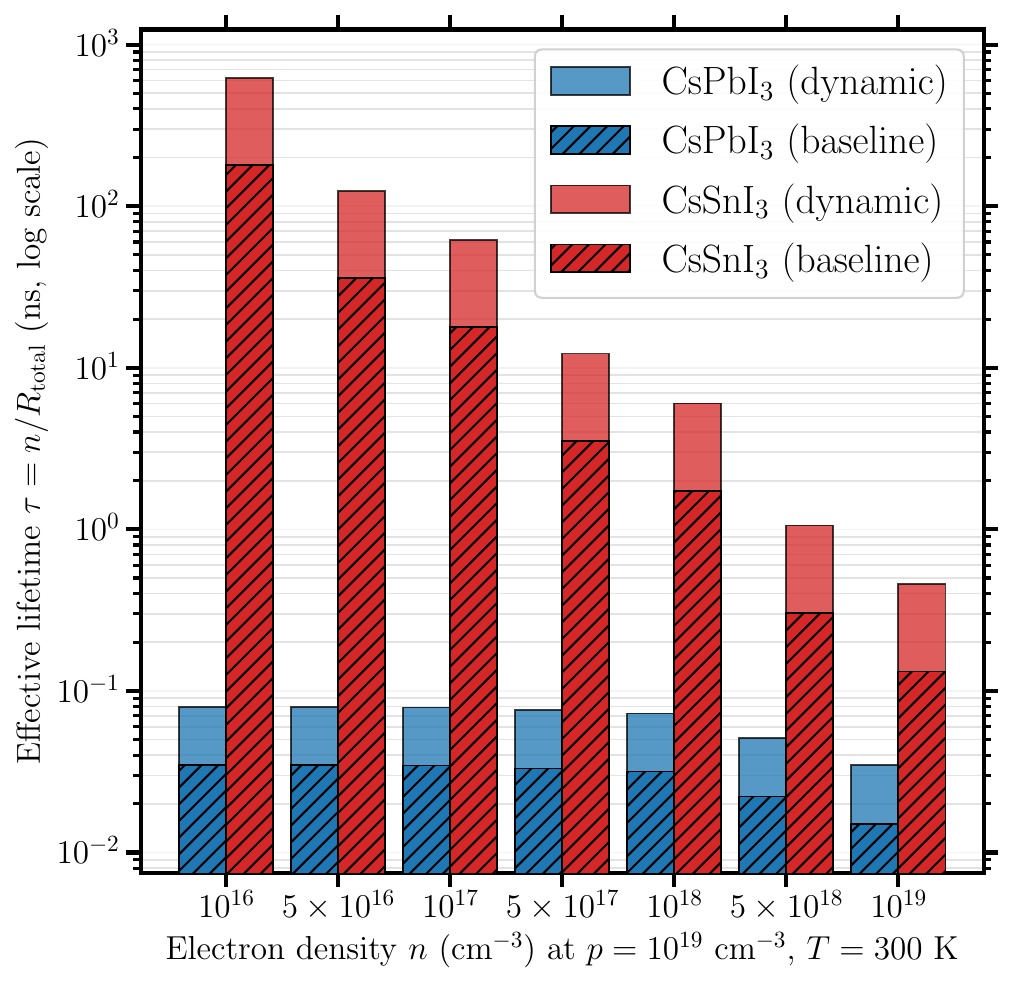}
\caption{%
Scan of effective Auger event lifetimes with electron density ($n$) with fixed higher hole density ($p$). For each $n$, CsPbI$_3$ (blue) and CsSnI$_3$ (red) lifetimes computed with dynamic screening (solid bars) are overlaid on the
corresponding 
baseline
lifetimes (hatched, semi-transparent) at the same
position. The $y$-axis is logarithmic.
}
\label{fig3}
\end{figure}
\paragraph{Material trends and implications for devices.}
Adopting the dynamically screened coefficient $C_{\mathrm{dyn}}=8.0\times10^{-28}\,\mathrm{cm^{6}\,s^{-1}}$ for $\gamma$-\ce{CsPbI3} at $300$\,K and a representative bimolecular radiative constant $B\simeq(1$-$2)\times10^{-10}\,\mathrm{cm^{3}\,s^{-1}}$, the density at which Auger and radiative lifetimes are equal is
\[
n_{\mathrm{crit}}=\frac{B}{C_{\mathrm{dyn}}}\approx (1.3\text{-}2.5)\times10^{17}\,\mathrm{cm^{-3}}.
\]
From the measured dynamic/baseline ratio at these conditions $C_{\mathrm{dyn}}/C_{\mathrm{base.}}=0.45$, we infer $C_{\mathrm{base}}\approx 1.78\times10^{-27}\,\mathrm{cm^{6}\,s^{-1}}$, which would place the crossover lower,
\[
n_{\mathrm{crit}}^{\mathrm{base.}}=\frac{B}{C_{\mathrm{base.}}}\approx (0.56\text{-}1.1)\times10^{17}\,\mathrm{cm^{-3}}.
\]
State-of-the-art PeLEDs and (opto)pumped perovskite lasers operate in high-injection regimes where carrier densities routinely approach $10^{18}\,\mathrm{cm^{-3}}$\cite{Sun2024APR,Qin2021TrChem,Dong2023eLight}. Thus, the $\sim\!55\%$ reduction in $C$ (dynamic vs.\ baseline) materially delays efficiency roll-off and lowers lasing thresholds in gain models where Auger dominates.

In Fig.~\ref{fig3}, the expected \emph{effective} lifetimes of an Auger event are shown for both \ce{CsSnI3} and \ce{CsPbI3}. While we already know that \ce{CsSnI3} is preferred over \ce{CsPbI3} for device applications where radiative recombination is desired, we now see that the change in evaluated lifetimes is much larger for \ce{CsSnI3} than \ce{CsPbI3}, which is favorable for device applications where low rate of Auger events is desired.

Because the dominant correction arises from the long-wavelength (head) part of the \emph{intrinsic} electronic response,\cite{Wilson2019APLM,Herz2018JPCL} embedding the material in external high-dielectric constant (high $\kappa$) encapsulants is expected to afford only limited additional leverage on the bulk Auger coefficient.
More effective routes could involve alloying on the $B$ or $X$ sites to achieve (i) band-structure detuning or strain to move the
CBM-CB resonance off of the Auger energy condition, and (ii) to reduce the interband overlaps via control over orbital-mixing control, thereby suppressing Auger matrix elements independently of dielectric screening.


In conclusion, we establish that a fully frequency-dependent treatment of dielectric screening is a \emph{prerequisite} for predictive modeling of Auger recombination in halide perovskites. By incorporating the fully dynamical treatment of the screened Coulomb interaction, our parameter-free calculations revise room-temperature coefficients by up to a factor of two, shifting predictions of LED operating regimes to higher, more favorable carrier densities. This quantitative renormalization -- originating from intrinsic many-body physics -- cannot be captured by conventional frequency-independent approaches. Beyond improved accuracy, this general framework identifies actionable materials-design levers in B-site composition and band alignment, and is readily transferable to other polar semiconductors where frequency-independent screening models are inadequate.

\vspace{0.5em}
\begin{acknowledgments}
U.S. thanks Johan Klarbring for valuable discussions. U.S. and
S.I.S. acknowledge support from the Swedish Research Council (VR) (2023-05247), the Swedish e-Science Center (SeRC), the ERC synergy grant (FASTCORR project 854843) and the Knut and Alice Wallenberg Foundation (Grant No. KAW 2019.0082). The computations were enabled by resources provided by the National Academic Infrastructure for Supercomputing in Sweden (NAISS), partially funded by the Swedish Research Council (2022-06725).
\end{acknowledgments}

Data availability. -- The numerical data used to generate the figures in this work are available at ~\cite{auger2025_data}.

\bibliographystyle{apsrev4-2}
\nocite{*}
\bibliography{main}

@article{Shen2018AEM,
  author  = {Shen, Jimmy-Xuan and Zhang, Xie and Das, Suvadip and Kioupakis, Emmanouil and Van de Walle, Chris G.},
  title   = {Unexpectedly Strong {Auger} Recombination in Halide {Perovskites}},
  journal = {Advanced Energy Materials},
  year    = {2018},
  volume  = {8},
  pages   = {1801027},
  doi     = {10.1002/aenm.201801027}
}

@article{Qin2021TrChem,
  author  = {Qin, Jiajun and Liu, Xiao-Ke and Yin, Chunyang and Gao, Feng},
  title   = {Carrier Dynamics and Evaluation of Lasing Actions in Halide {Perovskites}},
  journal = {Trends in Chemistry},
  year    = {2021},
  volume  = {3},
  number  = {1},
  pages   = {34--46},
  doi     = {10.1016/j.trechm.2020.10.010}
}

@article{Sun2024APR,
  author  = {Sun, Yuqi and Chen, Si and Huang, Jun-Yu and Wu, Yuh-Renn and Greenham, Neil C.},
  title   = {Device Physics of {Perovskite} Light-Emitting Diodes},
  journal = {Applied Physics Reviews},
  year    = {2024},
  volume  = {11},
  number  = {4},
  pages   = {041418},
  doi     = {10.1063/5.0228117}
}

@article{Zou2020ACSNano,
  author  = {Zou, Chen and Liu, Yun and Ginger, David S. and Lin, Lih Y.},
  title   = {Suppressing Efficiency Roll-Off at High Current Densities for Ultra-Bright Green {Perovskite} Light-Emitting Diodes},
  journal = {ACS Nano},
  year    = {2020},
  volume  = {14},
  pages   = {6076--6086},
  doi     = {10.1021/acsnano.0c01817}
}

@article{PazosOuton2018JPCL,
  author  = {Pazos-Out{\'o}n, Luis M. and Xiao, T. Patrick and Yablonovitch, Eli},
  title   = {Fundamental Efficiency Limit of Lead Iodide Perovskite Solar Cells},
  journal = {The Journal of Physical Chemistry Letters},
  year    = {2018},
  volume  = {9},
  number  = {7},
  pages   = {1703--1711},
  doi     = {10.1021/acs.jpclett.7b03054}
}

@article{Bowman2021ACSEnergy,
  author  = {Bowman, Alan R. and Lang, Felix and Chiang, Yu-Hsien and Jim{\'e}nez-Solano, Alberto and Frohna, Kyle and Eperon, Giles E. and Ruggeri, Edoardo and Abdi-Jalebi, Mojtaba and Anaya, Miguel and Lotsch, Bettina V. and Stranks, Samuel D.},
  title   = {Relaxed Current Matching Requirements in Highly Luminescent Perovskite Tandem Solar Cells and Their Fundamental Efficiency Limits},
  journal = {ACS Energy Letters},
  year    = {2021},
  volume  = {6},
  pages   = {612--620},
  doi     = {10.1021/acsenergylett.0c02481}
}

@article{Govoni2011PRB,
  author  = {Govoni, Marco and Marri, Ivan and Ossicini, Stefano},
  title   = {Auger recombination in Si and GaAs semiconductors: Ab initio results},
  journal = {Physical Review B},
  year    = {2011},
  volume  = {84},
  number  = {7},
  pages   = {075215},
  doi     = {10.1103/PhysRevB.84.075215}
}

@article{Steiauf2014ACSPho,
  author  = {Steiauf, Daniel and Kioupakis, Emmanouil and Van de Walle, Chris G.},
  title   = {Auger Recombination in GaAs from First Principles},
  journal = {ACS Photonics},
  year    = {2014},
  volume  = {1},
  number  = {8},
  pages   = {643--646},
  doi     = {10.1021/ph500119q}
}

@article{Zhang2020AEM,
  author  = {Zhang, Xie and Shen, Jimmy-Xuan and Van de Walle, Chris G.},
  title   = {First-Principles Simulation of Carrier Recombination Mechanisms in Halide Perovskites},
  journal = {Advanced Energy Materials},
  year    = {2020},
  volume  = {10},
  pages   = {1902830},
  doi     = {10.1002/aenm.201902830}
}

@article{Kioupakis2015PRB,
  author  = {Kioupakis, Emmanouil and Steiauf, Daniel and Rinke, Patrick and Delaney, Kris T. and Van de Walle, Chris G.},
  title   = {First-principles calculations of indirect {Auger} recombination in nitride semiconductors},
  journal = {Physical Review B},
  year    = {2015},
  volume  = {92},
  number  = {3},
  pages   = {035207},
  doi     = {10.1103/PhysRevB.92.035207}
}

@article{Herz2018JPCL,
  author  = {Herz, Laura M.},
  title   = {How Lattice Dynamics Moderate the Electronic Properties of Metal-Halide Perovskites},
  journal = {The Journal of Physical Chemistry Letters},
  year    = {2018},
  volume  = {9},
  number  = {24},
  pages   = {6853--6863},
  doi     = {10.1021/acs.jpclett.8b02811}
}

@article{Leveillee2019PRB,
  author  = {Leveillee, Joshua and Schleife, Andr{\'e}},
  title   = {Free-electron effects on the optical absorption of the hybrid perovskite CH$_3$NH$_3$PbI$_3$ from first principles},
  journal = {Physical Review B},
  year    = {2019},
  volume  = {100},
  number  = {3},
  pages   = {035205},
  doi     = {10.1103/PhysRevB.100.035205}
}

@article{Alymov2018PRB,
  author  = {Alymov, Georgy and Dmitriev, Alexey P. and Svintsov, Dmitry},
  title   = {Auger recombination in Dirac materials: A tangle of many-body effects},
  journal = {Physical Review B},
  year    = {2018},
  volume  = {97},
  number  = {20},
  pages   = {205411},
  doi     = {10.1103/PhysRevB.97.205411}
}

@article{Bushick2023PRL,
  author  = {Bushick, Kyle and Kioupakis, Emmanouil},
  title   = {Phonon-Assisted {Auger–Meitner} Recombination in Silicon from First Principles},
  journal = {Physical Review Letters},
  year    = {2023},
  volume  = {131},
  number  = {7},
  pages   = {076902},
  doi     = {10.1103/PhysRevLett.131.076902}
}

@article{Onida2002RMP,
  author  = {Onida, Giovanni and Reining, Lucia and Rubio, Angel},
  title   = {Electronic excitations: density-functional versus many-body Green's-function approaches},
  journal = {Reviews of Modern Physics},
  year    = {2002},
  volume  = {74},
  number  = {2},
  pages   = {601--659},
  doi     = {10.1103/RevModPhys.74.601}
}

@article{Wilson2019APLM,
  author  = {Wilson, Jacob N. and Frost, Jarvist M. and Wallace, Suzanne K. and Walsh, Aron},
  title   = {Dielectric and ferroic properties of metal halide perovskites},
  journal = {APL Materials},
  year    = {2019},
  volume  = {7},
  number  = {1},
  pages   = {010901},
  doi     = {10.1063/1.5079633}
}

@article{Dong2023eLight,
  author  = {Dong, H. and Ran, C. and Gao, W. and et al.},
  title   = {Metal Halide Perovskite for next-generation optoelectronics: progresses and prospects},
  journal = {eLight},
  year    = {2023},
  volume  = {3},
  pages   = {3},
  doi     = {10.1186/s43593-022-00033-z}
}

@article{Chen2016NatCommun,
  author  = {Chen, Y. and Yi, H. T. and Wu, X. and Haroldson, R. and Gartstein, Y. N. and Rodionov, Y. I. and Tikhonov, K. S. and Zakhidov, A. and Zhu, X.-Y. and Podzorov, V.},
  title   = {Extended carrier lifetimes and diffusion in hybrid perovskites revealed by Hall effect and photoconductivity measurements},
  journal = {Nature Communications},
  year    = {2016},
  volume  = {7},
  pages   = {12253},
  doi     = {10.1038/ncomms12253}
}

@article{Kresse1996PRB,
  author  = {Kresse, Georg and Furthm{\"u}ller, J{\"u}rgen},
  title   = {Efficient Iterative Schemes for \emph{Ab Initio} Total-Energy Calculations Using a Plane-Wave Basis Set},
  journal = {Physical Review B},
  year    = {1996},
  volume  = {54},
  number  = {16},
  pages   = {11169--11186},
  doi     = {10.1103/PhysRevB.54.11169}
}

@article{KresseFurthmueller1996CMS,
  author  = {Kresse, Georg and Furthm{\"u}ller, J{\"u}rgen},
  title   = {Efficiency of \emph{ab-initio} Total Energy Calculations for Metals and Semiconductors Using a Plane-Wave Basis Set},
  journal = {Computational Materials Science},
  year    = {1996},
  volume  = {6},
  number  = {1},
  pages   = {15--50},
  doi     = {10.1016/0927-0256(96)00008-0}
}

@article{KresseHafner1993PRB,
  author  = {Kresse, Georg and Hafner, J{\"o}rg},
  title   = {\emph{Ab Initio} Molecular Dynamics for Liquid Metals},
  journal = {Physical Review B},
  year    = {1993},
  volume  = {47},
  number  = {1},
  pages   = {558--561},
  doi     = {10.1103/PhysRevB.47.558}
}

@article{Blochl1994PAW,
  author  = {Bl{\"o}chl, Peter E.},
  title   = {Projector Augmented-Wave Method},
  journal = {Physical Review B},
  year    = {1994},
  volume  = {50},
  number  = {24},
  pages   = {17953--17979},
  doi     = {10.1103/PhysRevB.50.17953}
}

@article{KresseJoubert1999PAW,
  author  = {Kresse, Georg and Joubert, Daniel},
  title   = {From Ultrasoft Pseudopotentials to the Projector Augmented-Wave Method},
  journal = {Physical Review B},
  year    = {1999},
  volume  = {59},
  number  = {3},
  pages   = {1758--1775},
  doi     = {10.1103/PhysRevB.59.1758}
}

@article{Perdew2008PBEsol,
  author  = {Perdew, John P. and Ruzsinszky, Adrienn and Csonka, G\'abor I. and Vydrov, Oleg A. and Scuseria, Gustavo E. and Constantin, Lucian A. and Zhou, Xiaolan and Burke, Kieron},
  title   = {Restoring the Density-Gradient Expansion for Exchange in Solids and Surfaces},
  journal = {Physical Review Letters},
  year    = {2008},
  volume  = {100},
  number  = {13},
  pages   = {136406},
  doi     = {10.1103/PhysRevLett.100.136406}
}

@article{MonkhorstPack1976,
  author  = {Monkhorst, Hendrik J. and Pack, James D.},
  title   = {Special Points for Brillouin-Zone Integrations},
  journal = {Physical Review B},
  year    = {1976},
  volume  = {13},
  number  = {12},
  pages   = {5188--5192},
  doi     = {10.1103/PhysRevB.13.5188}
}

@article{Blochl1994Tetra,
  author  = {Bl{\"o}chl, Peter E. and Jepsen, Ole and Andersen, Ole K.},
  title   = {Improved Tetrahedron Method for Brillouin-Zone Integrations},
  journal = {Physical Review B},
  year    = {1994},
  volume  = {49},
  number  = {23},
  pages   = {16223--16233},
  doi     = {10.1103/PhysRevB.49.16223}
}

@article{MethfesselPaxton1989,
  author  = {Methfessel, Michael and Paxton, Anthony T.},
  title   = {High-Precision Sampling for Brillouin-Zone Integration in Metals},
  journal = {Physical Review B},
  year    = {1989},
  volume  = {40},
  number  = {6},
  pages   = {3616--3621},
  doi     = {10.1103/PhysRevB.40.3616}
}

@article{Hedin1965PR,
  author  = {Hedin, Lars},
  title   = {New Method for Calculating the One-Particle Green's Function with Application to the Electron-Gas Problem},
  journal = {Physical Review},
  year    = {1965},
  volume  = {139},
  number  = {3A},
  pages   = {A796--A823},
  doi     = {10.1103/PhysRev.139.A796}
}

@article{ShishkinKresse2006,
  author  = {Shishkin, M. and Kresse, Georg},
  title   = {Implementation and Performance of the Frequency-Dependent \(\mathrm{GW}\) Method within the PAW Framework},
  journal = {Physical Review B},
  year    = {2006},
  volume  = {74},
  number  = {3},
  pages   = {035101},
  doi     = {10.1103/PhysRevB.74.035101}
}

@article{Rojas1995PRL,
  author  = {Rojas, H.~N. and Godby, R.~W. and Needs, R.~J.},
  title   = {Space–Time Method for \emph{Ab Initio} Calculations of Self-Energies and Dielectric Response Functions of Solids},
  journal = {Physical Review Letters},
  year    = {1995},
  volume  = {74},
  number  = {10},
  pages   = {1827--1830},
  doi     = {10.1103/PhysRevLett.74.1827}
}

@article{Rieger1999CPC,
  author  = {Rieger, M.~M. and Steinbeck, L. and White, I.~D. and Rojas, H.~N. and Godby, R.~W.},
  title   = {The \(\mathrm{GW}\) Space–Time Method for the Self-Energy of Large Systems},
  journal = {Computer Physics Communications},
  year    = {1999},
  volume  = {117},
  number  = {3},
  pages   = {211--228},
  doi     = {10.1016/S0010-4655(98)00174-X}
}

@article{Steinbeck2000CPC,
  author  = {Steinbeck, L. and Rubio, A. and Reining, L. and Torrent, M. and White, I.~D. and Godby, R.~W.},
  title   = {Enhancements to the \(\mathrm{GW}\) Space–Time Method},
  journal = {Computer Physics Communications},
  year    = {2000},
  volume  = {125},
  number  = {1--3},
  pages   = {105--118},
  doi     = {10.1016/S0010-4655(99)00466-X}
}

@article{HybertsenLouie1986PRB,
  author  = {Hybertsen, Mark S. and Louie, Steven G.},
  title   = {Electron Correlation in Semiconductors and Insulators: Band Gaps and Quasiparticle Energies},
  journal = {Physical Review B},
  year    = {1986},
  volume  = {34},
  number  = {8},
  pages   = {5390--5413},
  doi     = {10.1103/PhysRevB.34.5390}
}

@article{VidbergSerene1977,
  author  = {Vidberg, H.~J. and Serene, J.~W.},
  title   = {Solving the Eliashberg Equations by Means of \(N\)-Point Pad{\'e} Approximants},
  journal = {Journal of Low Temperature Physics},
  year    = {1977},
  volume  = {29},
  number  = {3--4},
  pages   = {179--192},
  doi     = {10.1007/BF00655090}
}

@book{Baker1996,
  author    = {Baker, George A. and Graves-Morris, Peter},
  title     = {Pad{\'e} Approximants},
  edition   = {2},
  publisher = {Cambridge University Press},
  year      = {1996},
  doi       = {10.1017/CBO9780511530074}
}

@article{Golze2019FrontChem,
  author  = {Golze, Dorothea and Dvorak, Marc and Rinke, Patrick},
  title   = {The \(\mathrm{GW}\) Compendium: A Practical Guide to Theoretical Photoemission Spectroscopy},
  journal = {Frontiers in Chemistry},
  year    = {2019},
  volume  = {7},
  pages   = {377},
  doi     = {10.3389/fchem.2019.00377}
}

@article{Adler1962PR,
  author  = {Adler, Stephen L.},
  title   = {Quantum Theory of the Dielectric Constant in Real Solids},
  journal = {Physical Review},
  year    = {1962},
  volume  = {126},
  number  = {2},
  pages   = {413--420},
  doi     = {10.1103/PhysRev.126.413}
}

@article{Wiser1963PR,
  author  = {Wiser, Nathan},
  title   = {Dielectric Constant with Local Field Effects Included},
  journal = {Physical Review},
  year    = {1963},
  volume  = {129},
  number  = {1},
  pages   = {62--69},
  doi     = {10.1103/PhysRev.129.62}
}

@book{Sze2006,
  author    = {Sze, Simon M. and Ng, Kwok K.},
  title     = {Physics of Semiconductor Devices},
  edition   = {3},
  publisher = {Wiley},
  year      = {2006},
  doi       = {10.1002/0470068329}
}

@book{Mahan2000,
  author    = {Mahan, Gerald D.},
  title     = {Many-Particle Physics},
  edition   = {3},
  publisher = {Kluwer/Plenum},
  year      = {2000},
  doi       = {10.1007/978-1-4757-5714-9}
}

@book{HaugKoch2009,
  author    = {Haug, Hartmut and Koch, Stephan W.},
  title     = {Quantum Theory of the Optical and Electronic Properties of Semiconductors},
  edition   = {5},
  publisher = {World Scientific},
  year      = {2009},
  doi       = {10.1142/7184}
}

@book{RobertCasella2004,
  author    = {Robert, Christian P. and Casella, George},
  title     = {Monte Carlo Statistical Methods},
  edition   = {2},
  publisher = {Springer},
  year      = {2004},
  doi       = {10.1007/978-1-4757-4145-2}
}

@book{Chopin2020,
  author    = {Nicolas Chopin and Omiros Papaspiliopoulos},
  title     = {An Introduction to Sequential Monte Carlo},
  series    = {Springer Series in Statistics},
  publisher = {Springer},
  year      = {2020},
  isbn      = {978-3-030-47844-5},
  doi       = {10.1007/978-3-030-47845-2},
  url       = {https://doi.org/10.1007/978-3-030-47845-2}
}

@article{MarzariVanderbilt1997,
  author  = {Marzari, Nicola and Vanderbilt, David},
  title   = {Maximally Localized Generalized Wannier Functions for Composite Energy Bands},
  journal = {Physical Review B},
  year    = {1997},
  volume  = {56},
  number  = {20},
  pages   = {12847--12865},
  doi     = {10.1103/PhysRevB.56.12847}
}

@article{SouzaMarzariVanderbilt2001,
  author  = {Souza, Ivo and Marzari, Nicola and Vanderbilt, David},
  title   = {Maximally Localized Wannier Functions for Entangled Energy Bands},
  journal = {Physical Review B},
  year    = {2001},
  volume  = {65},
  pages   = {035109},
  doi     = {10.1103/PhysRevB.65.035109}
}

@article{Pizzi2020,
  author  = {Pizzi, Giovanni and Vitale, Valerio and Arita, Ryotaro and Bl{\"u}gel, Stefan and Freimuth, Frank and et al.},
  title   = {Wannier90 as a Community Code: New Features and Applications},
  journal = {Journal of Physics: Condensed Matter},
  year    = {2020},
  volume  = {32},
  number  = {16},
  pages   = {165902},
  doi     = {10.1088/1361-648X/ab51ff}
}

@article{Giustino2017RMP,
  author  = {Giustino, Feliciano},
  title   = {Electron-Phonon Interactions from First Principles},
  journal = {Reviews of Modern Physics},
  year    = {2017},
  volume  = {89},
  pages   = {015003},
  doi     = {10.1103/RevModPhys.89.015003}
}

@article{GygiBaldereschi1986,
  author  = {Gygi, Fran{\c c}ois and Baldereschi, Alfonso},
  title   = {Self-Consistent Hartree–Fock and Screened-Exchange Calculations in Solids: Application to Silicon},
  journal = {Physical Review B},
  year    = {1986},
  volume  = {34},
  number  = {6},
  pages   = {4405--4408},
  doi     = {10.1103/PhysRevB.34.4405}
}

@book{Lucarini2005,
  author    = {Lucarini, Valerio and Saarinen, J.~J. and Peiponen, K.-E. and Vartiainen, E.~M.},
  title     = {Kramers–Kronig Relations in Optical Materials Research},
  publisher = {Springer},
  year      = {2005},
  doi       = {10.1007/b138913}
}

@article{Yuan2023,
	author = {Yuan, Fanglong and Folpini, Giulia and Liu, Tianjun and Singh, Utkarsh and Treglia, Antonella and Lim, Jia Wei Melvin and Klarbring, Johan and Simak, Sergei I. and Abrikosov, Igor A. and Sum, Tze Chien and Petrozza, Annamaria and Gao, Feng},
	date = {2024/02/01},
	doi = {10.1038/s41566-023-01351-5},
	id = {Yuan2024},
	isbn = {1749-4893},
	journal = {Nature Photonics},
	number = {2},
	pages = {170--176},
	title = {Bright and stable near-infrared lead-free perovskite light-emitting diodes},
	url = {https://doi.org/10.1038/s41566-023-01351-5},
	volume = {18},
	year = {2024}}

@misc{auger2025_data,
  author       = {Utkarsh Singh and Sergei I. Simak},
  title        = {Data for: ``{D}ynamic {S}creening {E}ffects on {A}uger
                  {R}ecombination in {M}etal-{H}alide {P}erovskites''},
  howpublished = {\url{https://github.com/utksi/auger2025}},
  year         = {2025},
  note         = {GitHub repository, commit a1e2e6d, accessed 8 December 2025}
}

\end{document}


\title{Supplementary Information for:\\
Dynamic Screening Effects on Auger Recombination in Metal-Halide Perovskites}

\author{Utkarsh Singh}
\email{utkarsh.singh[at]liu.se}
\affiliation{%
  Theoretical Physics Division, Department of Physics, Chemistry, and Biology (IFM),\\
  Link\"opings Universitet, SE-581 83 Link\"oping, Sweden
}

\author{Sergei I. Simak}
\affiliation{%
  Theoretical Physics Division, Department of Physics, Chemistry, and Biology (IFM),\\
  Link\"opings Universitet, SE-581 83 Link\"oping, Sweden
}
\affiliation{%
  Department of Physics and Astronomy, Uppsala University, SE-75120 Uppsala, Sweden
}

\maketitle

\tableofcontents

\clearpage


\section*{S1. Electronic structure calculations}

\paragraph*{DFT setup.}
All ground-state calculations and ionic relaxations were performed with the plane-wave PAW code VASP \cite{Kresse1996PRB,KresseFurthmueller1996CMS,KresseHafner1993PRB} using the projector-augmented-wave method \cite{Blochl1994PAW,KresseJoubert1999PAW}. Unless stated otherwise we employed the PBEsol exchange-correlation functional for accurate equilibrium volumes of dense solids \cite{Perdew2008PBEsol}. Brillouin-zone integrals used $\Gamma$-centered k-point density of 0.05 \AA$^{-1}$ \cite{MonkhorstPack1976}; for insulating states total energies and densities of states were evaluated using the tetrahedron method with Bl\"ochl corrections \cite{Blochl1994Tetra}. Structures were relaxed to tight thresholds (forces below typical $10^{-4}$\,eV/\AA\ and total-energy changes $\leq 10^{-7}$\,eV). For heavy-element halide perovskites, spin-orbit coupling (SOC) was included in single-point calculations that define the \(G_0W_0\) starting point.

\paragraph*{Single-shot \(G_0W_0\) and dielectric response.}
Quasiparticle corrections were evaluated within single-shot \(G_0W_0\) on top of the converged DFT(+SOC) reference, following Hedin’s \(GW\) formalism \cite{Hedin1965PR}. To reduce scaling, the polarizability and screened interaction were computed on the imaginary time/frequency axis using the real-space “space-time” scheme \cite{Rojas1995PRL,Rieger1999CPC,Steinbeck2000CPC}, within the RPA and PAW framework (representative implementation details in \cite{ShishkinKresse2006}). 
Analytic continuation to the real axis used a \emph{passive} (non‑negative) pole expansion fitted to the Matsubara data; we cross‑checked with a Thiele/Pad\'e rational approximant \cite{VidbergSerene1977,Baker1996}. Kramers-Kronig consistency checks were applied to the continued spectra.
Wannier-based interpolation used elsewhere in this S.I. is based on Maximally Localized Wannier Functions (MLWFs) and disentanglement \cite{MarzariVanderbilt1997,SouzaMarzariVanderbilt2001,Pizzi2020}.

\section*{S2. Notation and Brillouin-zone conventions}

We consider Bloch eigenstates $\ket{n\mathbf{k}}$ with band index $n$ and crystal
momentum $\mathbf{k}$ in the first Brillouin zone (BZ), eigenvalues
$\varepsilon_{n\mathbf{k}}$, and cell-periodic parts $u_{n\mathbf{k}}(\mathbf{r})$.
The real-space primitive cell volume is $\Omega$ and the BZ volume is
$V_{\mathrm{BZ}}=(2\pi)^3/\Omega$.

Sums over momenta are related to continuum integrals as
$\sum_{\mathbf{k}} \to \frac{N_k}{V_{\mathrm{BZ}}} \int_{\mathrm{BZ}} d^3k$, where
$N_k$ is the number of wavevectors on the uniform mesh used to
interpolate all quantities.

Under steady-state nonequilibrium (e.g., LED pumping) we adopt
separate quasi-Fermi levels for the conduction and valence manifolds,
$\mu_c$ and $\mu_v$, and temperature $T$. The Fermi-Dirac occupations are
\begin{equation}
f_c(\epsilon) = \frac{1}{1+e^{(\epsilon-\mu_c)/k_BT}},\qquad
f_v(\epsilon) = \frac{1}{1+e^{(\epsilon-\mu_v)/k_BT}},
\end{equation}
with carrier densities (per unit volume)
\begin{equation}
n = \frac{g_s}{\Omega N_k}\sum_{\mathbf{k},\,n\in\mathrm{CB}} f_c(\epsilon_{n\mathbf{k}}),\qquad
p = \frac{g_s}{\Omega N_k}\sum_{\mathbf{k},\,v\in\mathrm{VB}} \bigl[1-f_v(\epsilon_{v\mathbf{k}})\bigr].
\label{eq:density-defs}
\end{equation}
where $g_s$ is the spin degeneracy factor, CB and VB denote conduction and valence bands, respectively.
In practice, Eqs.~(\ref{eq:density-defs}) are inverted numerically to obtain
$\mu_c$ and $\mu_v$ for targeted $(n,p,T)$. The quasi-Fermi-level description is standard in semiconductor physics \cite{Sze2006}.

\vspace{0.25em}
\noindent\textit{Phase-space averaging notation.}
The angle brackets $\langle\cdots\rangle_{\mathcal{Q}}$ denote a weighted average over the sampled configuration space, with explicit Monte Carlo estimators given in Sec.~S9. For direct Auger, $\mathcal{Q}$ represents momentum- and energy-conserving quadruplets; for phonon-assisted processes, $\mathcal{Q}_5$ includes the additional phonon mode indices.

\vspace{0.25em}
\noindent\textit{Energy-conservation kernel.}
The energy delta in Fermi’s golden rule is represented numerically by a normalized
kernel $\delta_\sigma(\Delta E)$:
\begin{equation}
\delta_\sigma^{\mathrm{(gauss)}}(x)=\frac{1}{\sqrt{2\pi}\sigma}
e^{-x^2/2\sigma^2}
\quad\text{or}\quad
\delta_\sigma^{\mathrm{(tophat)}}(x)=\frac{\Theta(\sigma-|x|)}{2\sigma}.
\label{eq:delta}
\end{equation}
We use $\sigma=5~\mathrm{meV}$ with the gaussian kernel unless noted otherwise. Both choices are unit-normalized.


\section*{S3. Screened Coulomb interaction and dynamical screening}

We evaluate the $G=G'=0$ component (“head”) of the screened Coulomb interaction,
\begin{equation}
W_{00}(\mathbf q,\omega)
= \sum_{G'} \varepsilon^{-1}_{0G'}(\mathbf q,\omega)\,v(\mathbf q+\mathbf G'),
\label{eq:W00}
\end{equation}
where $v(\mathbf q)=4\pi/|\mathbf q|^2$ (Gaussian units) is the bare Coulomb interaction, $G$ and $G'$ are reciprocal lattice vectors, and $\varepsilon^{-1}_{GG'}$ is the inverse microscopic dielectric matrix within GW computed in the random phase approximation (RPA)
\cite{Hedin1965PR,Adler1962PR,Wiser1963PR,HybertsenLouie1986PRB}.

Defining the macroscopic dielectric function by
$\varepsilon_M^{-1}(\mathbf q,\omega)\equiv \varepsilon^{-1}_{00}(\mathbf q,\omega)$,
the long-wavelength (small-$|\mathbf q|$) reduction reads
\[
  W_{00}(\mathbf q,\omega)\simeq \frac{v(\mathbf q)}{\varepsilon_M(\mathbf q,\omega)}.
\]

We compute $\varepsilon^{-1}_{00}(\mathbf q,i\xi)$ 
on the imaginary axis (where $\xi$ denotes the Matsubara frequency) within the low-scaling $G_0W_0$ (imaginary time/frequency) framework
\cite{Rojas1995PRL,Rieger1999CPC,Steinbeck2000CPC}, 
and analytically continue to the real axis $\omega{+}i\eta$ (where $\eta$ is an infinitesimal positive broadening)
using a \emph{passive} (non-negative) pole expansion fitted to the
Matsubara data; see Sec.~S11 of the S.I. for passivity enforcement and Kramers-Kronig
consistency checks. As a cross-check only, we also perform Thiele/Pad\'e rational
continuations \cite{VidbergSerene1977,Baker1996}. Continuation errors from
Kramers-Kronig sum rules are below $2\%$ \cite{Lucarini2005}.

\paragraph*{Notation.}
In practice we work with the head-only screened interaction. By
$W_{00}^{\mathrm{ab\ initio}}(\mathbf q,\omega)$ we mean the microscopic head
obtained from the inversion of the full dielectric matrix (so that local-field
effects enter implicitly through the inversion) followed by multiplication by
the bare head potential:
\[
  W_{00}^{\mathrm{ab\ initio}}(\mathbf q,\omega)\;\equiv\;
  \varepsilon^{-1}_{00}(\mathbf q,\omega)\,v(\mathbf q),
\]
with $v(\mathbf q)=4\pi/|\mathbf q|^2$ (Gaussian units). For $|\mathbf q|\!\to\!0$
we use the macroscopic reduction $v(\mathbf q)/\varepsilon_M(\omega)$; the practical
$\Gamma$-cell average and the final real-axis assembly are given in Sec.~S11.

In what follows, “dynamic” screening means evaluating $W_{00}(\mathbf q,\omega)$ at the bosonic frequencies associated with the Coulomb legs. The \emph{frequency-independent baseline} uses $W_{00}(\mathbf q,\omega\!=\!0)$; in the long-wavelength head ($G{=}G'{=}0$, $q\!\to\!0$) this reduces to $v(\mathbf q)/\varepsilon_\infty$, where $\varepsilon_\infty$ denotes the ion-clamped (electronic) macroscopic dielectric constant.

Throughout the S.I. we refer to this as the \emph{baseline} calculation.



\section*{S4. Small- and large-\texorpdfstring{$\mathbf{q}$}{q} treatment and the \texorpdfstring{$q_c$}{qc} partition}

Local-field effects are negligible for $|\mathbf q|\!\ll\!|\mathbf G|$, and $W_{00}$ reduces to the macroscopic form $v(\mathbf q)/\varepsilon_M(\omega)$ with $\varepsilon_M(\omega)\equiv \varepsilon_M(q\!\to\!0,\omega)$ (microscopic vs.\ macroscopic screening as in \cite{Adler1962PR,Wiser1963PR,HybertsenLouie1986PRB}). 

We therefore define the assembled head used in the calculations by a $q$‑space partition at radius $q_c$:
\begin{equation}
W_{00}(\mathbf q,\omega)=
\begin{cases}
\displaystyle \frac{v(\mathbf q)}{\varepsilon_M(\omega)}, & |\mathbf q|\le q_c,\\[0.75em]
\displaystyle W_{00}^{\mathrm{ab\ initio}}(\mathbf q,\omega), & |\mathbf q|>q_c,
\end{cases}
\label{eq:qc-split}
\end{equation}
with $v(\mathbf q)=4\pi/|\mathbf q|^{2}$ (Gaussian units).
The cutoff $q_c$ is set by the BZ discretization (Sec. S11).
The explicit real-axis assembly for both branches (including the $\Gamma$-cell average) is given in Sec.~S11.



\section*{S5. Direct Auger recombination (eeh and hhe)}

Consider a direct $eeh$ event where two electrons (1,2) annihilate a hole (3), promoting
electron (4). The coherent golden-rule rate density is
\begin{align}
R_{eeh}^{\mathrm{dir}}
&=\frac{2\pi}{\hbar}\,\frac{V_{\mathrm{BZ}}^{3}}{\Omega}\,
\Big\langle \delta_{\sigma}\!\bigl(\Delta E\bigr)\;
P_{eeh}\;\bigl|M_{eeh}\bigr|^{2}\Big\rangle_{\mathcal{Q}},
\label{eq:Reeh-master}
\\
\Delta E&=\epsilon_{1}+\epsilon_{2}-\epsilon_{3}-\epsilon_{4},\qquad
P_{eeh}=f_c(\epsilon_{1})\,f_c(\epsilon_{2})\,
\bigl[1-f_v(\epsilon_{3})\bigr]\,\bigl[1-f_c(\epsilon_{4})\bigr],
\end{align}
where $\langle\cdots\rangle_{\mathcal{Q}}$ denotes a weighted average over the set of momentum- and energy-conserving quadruplets (explicit estimator in Sec.~S9),
and the coherent (antisymmetrized) amplitude is
\begin{align}
M_{eeh} &= M_{eeh}^{\mathrm{dir}} - M_{eeh}^{\mathrm{exc}},
\\
M_{eeh}^{\mathrm{dir}}
&= W_{00}\!\bigl(\mathbf{q}_{41},\omega_{41}\bigr)\;
\underbrace{\braket{u_{4\mathbf{k}_4}}{u_{1\mathbf{k}_1}}}_{O_{41}}\,
\underbrace{\braket{u_{2\mathbf{k}_2}}{u_{3\mathbf{k}_3}}}_{O_{23}},
\label{eq:Mdir-eeh}
\\
M_{eeh}^{\mathrm{exc}}
&= W_{00}\!\bigl(\mathbf{q}_{31},\omega_{31}\bigr)\;
\underbrace{\braket{u_{3\mathbf{k}_3}}{u_{1\mathbf{k}_1}}}_{O_{31}}\,
\underbrace{\braket{u_{2\mathbf{k}_2}}{u_{4\mathbf{k}_4}}}_{O_{24}}.
\label{eq:Mexc-eeh}
\end{align}
The exchanged channel comes with a minus sign by fermionic antisymmetry. The
momentum transfers are $\mathbf{q}_{41}=\mathbf{k}_4-\mathbf{k}_1+\mathbf{G}_{41}$
and $\mathbf{q}_{31}=\mathbf{k}_3-\mathbf{k}_1+\mathbf{G}_{31}$ (Umklapp allowed),
and the bosonic frequency arguments are set by the corresponding electronic
energy transfers, e.g.
$\hbar\omega_{41}=\epsilon_{4}-\epsilon_{1}$ and
$\hbar\omega_{31}=\epsilon_{3}-\epsilon_{1}$.
The overlaps between cell-periodic Bloch functions are defined as
$O_{ab} \equiv \braket{u_{a\mathbf{k}_a}}{u_{b\mathbf{k}_b}}$
and are computed from Wannier-interpolated wavefunctions (see Sec.~S10).

The $hhe$ channel is obtained by interchanging $c\!\leftrightarrow\! v$ and
relabelling:
\begin{equation}
P_{hhe}=\bigl[1-f_v(\epsilon_{1})\bigr]\bigl[1-f_v(\epsilon_{2})\bigr]
f_c(\epsilon_{3})\,f_v(\epsilon_{4}),
\end{equation}
with the same structure for $M_{hhe}$ as Eqs.~(\ref{eq:Mdir-eeh})-(\ref{eq:Mexc-eeh})
but with overlaps consistent with the $(v,v;c,v)$ slot roles.

\section*{S6. Phonon-assisted Auger recombination}

In the phonon-assisted ($\nu,\mathbf{q}$) channel, a single phonon of branch $\nu$
and momentum $\mathbf{q}$ is absorbed $(s=+1)$ or emitted $(s=-1)$.
The coherent golden-rule rate density reads
\begin{align}
R_{eeh}^{\mathrm{ph}}
&=\frac{2\pi}{\hbar}\,\frac{V_{\mathrm{BZ}}^{4}}{\Omega\,N_q}\,
\Big\langle \delta_{\sigma}\!\bigl(\Delta E - s\,\hbar\Omega_{\mathbf{q}\nu}\bigr)\;
P_{eeh}\;\mathcal{S}_s(\Omega_{\mathbf{q}\nu},T)\;
\bigl|M_{eeh}^{\mathrm{ph}}(\nu,\mathbf{q})\bigr|^{2}\Big\rangle_{\mathcal{Q}_5},
\label{eq:Rph-master}
\end{align}
with $\mathcal{Q}_5$ denoting sampled quintuplets $(\mathbf{k}_1,\mathbf{k}_2,\mathbf{k}_3,\mathbf{k}_4;\mathbf{q},\nu,s)$,
$N_q$ the number of $\mathbf{q}$ points entering the electron-phonon database,
and the Bose stimulation factor
\begin{equation}
\mathcal{S}_s(\Omega,T)=
\begin{cases}
n_B(\Omega,T)=\bigl[e^{\hbar\Omega/k_BT}-1\bigr]^{-1},& s=+1\ \text{(abs.)},\\[0.25em]
1+n_B(\Omega,T), & s=-1\ \text{(em.)}.\\
\end{cases}
\label{eq:bose}
\end{equation}
In second-order perturbation theory the coherent amplitude is a sum over virtual
intermediate states $\{\ket{m}\}$,
\begin{equation}
M_{eeh}^{\mathrm{ph}}(\nu,\mathbf{q})=
\sum_{m}\frac{\bra{f} \hat{H}_C \ket{m}\,\bra{m}\hat{H}_{e\text{-}ph}^{\nu}(\mathbf{q})\ket{i}}
{E_i-E_m+s\hbar\Omega_{\mathbf{q}\nu}+ i0^+}
\quad \text{(direct$-$exchange)},
\label{eq:Mph-formal}
\end{equation}
where $\hat{H}_C$ is the screened Coulomb interaction and
$\hat{H}_{e\text{-}ph}^{\nu}$ the linear electron-phonon coupling.
In the present implementation we evaluate the \emph{coherent} Coulomb factor in
the same way as for direct Auger, Eqs.~(\ref{eq:Mdir-eeh})-(\ref{eq:Mexc-eeh}), and
multiply $|M|^2$ by the Frobenius norm $\|g\|^2$ of the mode‑resolved
electron-phonon matrix elements $g_{mn\nu}(\mathbf{k},\mathbf{q})$ attached to each sampled quintuplet. This
“diagonal” treatment neglects phase correlations among different virtual states,
a good approximation when the e-ph couplings entering a given quintuplet are
localized in band space and the electronic denominators vary slowly over the
dominant contribution region.
The $hhe$ phonon-assisted channel has the same structure with the corresponding
slot roles. Equation~(\ref{eq:Rph-master}) treats absorption and emission on equal
footing via $s=\pm1$.


\section*{S7. Baseline vs dynamic screening}

“Dynamic” results evaluate $W_{00}(\mathbf q,\omega)$ at the bosonic frequencies associated with the Coulomb legs: $\hbar\omega_{41}=\varepsilon_4-\varepsilon_1$ (direct) and $\hbar\omega_{31}=\varepsilon_3-\varepsilon_1$ (exchange) in Eqs.~(\ref{eq:Mdir-eeh})-(\ref{eq:Mexc-eeh}), with the $q$‑sector partition of Eq.~(\ref{eq:qc-split}). The baseline screening results 
use $W_{00}(\mathbf q)$, which in the long‑wavelength sector ($|\mathbf q|\le q_c$), reduces to $v(\mathbf q)/\varepsilon_\infty$, where $\varepsilon_\infty\equiv \varepsilon_M^{\mathrm{elec}}(q\!\to\!0,\omega{=}0)$.




\section*{S8. Auger coefficients}

We report the conventional coefficients
\begin{equation}
C_n(T;n,p)=\frac{R_{eeh}(T;n,p)}{n^2 p},\qquad
C_p(T;n,p)=\frac{R_{hhe}(T;n,p)}{p^2 n}.
\label{eq:CnCp}
\end{equation}
with $R_{eeh}=R_{eeh}^{\mathrm{dir}}+R_{eeh}^{\mathrm{ph}}$ and
$R_{hhe}=R_{hhe}^{\mathrm{dir}}+R_{hhe}^{\mathrm{ph}}$, all as rate densities
($\mathrm{cm^{-3}\,s^{-1}}$). No single “total $C$” is defined unless one
imposes a specific relation between $n$ and $p$ (e.g.\ $n=p$).


\section*{S9. Monte Carlo estimators and importance sampling}

The $k$-space constrained integrals in Eqs.~(\ref{eq:Reeh-master})
and (\ref{eq:Rph-master}) are evaluated by importance sampling of
quadruplets $\mathcal{Q}$ for direct and quintuplets $\mathcal{Q}_5$ for
phonon-assisted processes. Each hit carries a positive weight $w_j$ that
encodes the sampler’s proposal density and any symmetry factors.
The explicit Monte Carlo estimators are
\begin{align}
R_{eeh}^{\mathrm{dir}}
&=\frac{2\pi}{\hbar}\,\frac{V_{\mathrm{BZ}}^{3}}{\Omega}\,
\frac{\sum_{j\in\mathcal{Q}} w_j\,\delta_\sigma(\Delta E_j)\,P_{eeh,j}\,|M_{eeh,j}|^2}
{\sum_{j\in\mathcal{Q}} w_j},
\\
R_{eeh}^{\mathrm{ph}}
&=\frac{2\pi}{\hbar}\,\frac{V_{\mathrm{BZ}}^{4}}{\Omega\,N_q}\,
\frac{\sum_{j\in\mathcal{Q}_5} w_j\,\delta_\sigma(\Delta E_j-s_j\hbar\Omega_j)\,P_{eeh,j}\,
\mathcal{S}_{s_j}(\Omega_j,T)\,|M^{\mathrm{ph}}_{eeh,j}|^2}
{\sum_{j\in\mathcal{Q}_5} w_j},
\end{align}
and analogously for $hhe$. 
All microscopic quantities (overlaps, electron-phonon matrix elements $g$, phonon frequencies) are obtained by maximally localized Wannier function (MLWF) interpolation onto the sampler's
dense meshes.
The overall prefactors $(2\pi/\hbar)\,V_{\mathrm{BZ}}^3/\Omega$
and $(2\pi/\hbar)\,V_{\mathrm{BZ}}^4/(\Omega N_q)$ arise from
the dimensional reduction imposed by momentum conservation.
For Monte Carlo methodology details and design references, see \cite{RobertCasella2004,Chopin2020}.


\section*{S10. Overlaps, channel geometry, and \texorpdfstring{$W$}{W}-dataset mapping}

The overlaps in Eqs.~(\ref{eq:Mdir-eeh})-(\ref{eq:Mexc-eeh}) are
\begin{equation}
O_{ab} \equiv \braket{u_{a\mathbf{k}_a}}{u_{b\mathbf{k}_b}}=
\int_{\Omega} u_{a\mathbf{k}_a}^{*}(\mathbf{r})\,u_{b\mathbf{k}_b}(\mathbf{r})\,d^3r,
\end{equation}
computed from the column-projected unitary rotations that transform
the Wannier gauge to the band gauge.
We employ standard MLWF/disentanglement workflows for robust interpolation \cite{MarzariVanderbilt1997,SouzaMarzariVanderbilt2001,Pizzi2020}.
In the direct channel the physically
correct transferred momentum is $\mathbf{q}_{41}=\mathbf{k}_4-\mathbf{k}_1+ \mathbf{G}_{41}$
(and analogously $\mathbf{q}_{31}$ in exchange). Precomputed datasets with
$W$ arrays are mapped to the relevant quartet or quintuplet by matching their associated $\mathbf{q}$s
against $\mathbf{q}_{41}$ and $\mathbf{q}_{31}$.
\section*{S11. Imaginary-axis \texorpdfstring{$GW$}{GW} and analytic continuation}
\paragraph*{Matsubara grid.}
Unless stated otherwise we use $N_\xi=66$ points with a denser sampling near
$\xi{=}0$ and a high-frequency cutoff $\xi_{\max}\!\approx\!60$\,eV; the same
$\{\xi_j\}$ is used for all $\mathbf{q}$.

\paragraph*{$\Gamma$-cell average for small $q$.}
To regularize the $1/q^2$ head and eliminate special-casing downstream, we define a
$\Gamma$-sphere of radius $q_c$ whose volume equals the irreducible Brillouin zone (IBZ) weight of $\Gamma$,
\begin{equation}
  \frac{4\pi}{3}\,q_c^{3} \;=\; w_\Gamma\,V_{\mathrm{BZ}}
  \;=\; w_\Gamma\,\frac{(2\pi)^3}{\Omega}
  \quad\Rightarrow\quad
  q_c \;=\; \Bigl(\frac{6\pi^2\,w_\Gamma}{\Omega}\Bigr)^{1/3}.
  \label{eq:qc-from-weight}
\end{equation}
Over this ball the average of $1/q^2$ is $\langle 1/q^2\rangle = 3/q_c^2$, so the average of
$v(\mathbf q)=4\pi/|\mathbf q|^2$ (Gaussian units) is
$\langle v\rangle_{|\mathbf q|\le q_c}=12\pi/q_c^2$. For $|q| \le q_c$ the macroscopic form reproduces the microscopic head to within a few percent, justifying the partition in Eq. S5. This $\Gamma$-cell average follows the standard treatment of the Coulomb singularity in periodic Brillouin-zone integrations \cite{GygiBaldereschi1986}.
We store a finite central-cell value for all $|\mathbf q|\le q_c$:
\begin{equation}
  W_{00}^{\mathrm{cell}}(\omega{+}i\eta)
  \;=\; \varepsilon^{-1}_{M}(\omega{+}i\eta)\,\frac{12\pi}{q_c^{2}}.
  \label{eq:gamma-cell}
\end{equation}
For $|\mathbf q|>q_c$ we use the head-only microscopic form. The final assembled head on the real axis is
\begin{equation}
  W_{00}(\mathbf{q},\omega{+}i\eta) \;=\;
  \begin{cases}
    \displaystyle \varepsilon^{-1}_{M}(\omega{+}i\eta)\,\frac{12\pi}{q_c^{2}}, & |\mathbf{q}|\le q_c,\\[0.6em]
    \displaystyle \varepsilon^{-1}_{00}(\mathbf{q},\omega{+}i\eta)\,\frac{4\pi}{|\mathbf{q}|^{2}}, & |\mathbf{q}|>q_c.
  \end{cases}
  \label{eq:W-assembly}
\end{equation}

\noindent Equivalently, on the $|\mathbf q|>q_c$ branch,
$W_{00}(\mathbf q,\omega{+}i\eta)=W_{00}^{\mathrm{ab\ initio}}(\mathbf q,\omega{+}i\eta)$
by definition.

We fit the imaginary-axis data for each $\mathbf{q}$ by a passive pole expansion
(nonnegative strengths, real pole frequencies):
\begin{align}
  \varepsilon^{-1}_{00}(\mathbf{q},i\xi)
  &\approx 1 - \sum_{j=1}^{N_p}\frac{B_j(\mathbf{q})}{1+\xi^{2}/\Omega_j^{2}},
  \qquad B_j(\mathbf{q})\ge 0,\ \ \Omega_j>0,
  \label{eq:nnls-imag}\\
  \varepsilon^{-1}_{00}(\mathbf{q},\omega{+}i\eta)
  &\approx 1 - \sum_{j=1}^{N_p}\frac{B_j(\mathbf{q})}{1-(\omega{+}i\eta)^{2}/\Omega_j^{2}}.
  \label{eq:nnls-real}
\end{align}
The parameters $\{B_j,\Omega_j\}$ are obtained by nonnegative least squares on the
imaginary axis using a logarithmically spaced set of trial $\{\Omega_j\}$,
with cross-checks against a Thiele/Pad\'e approximant \cite{VidbergSerene1977,Baker1996}.
The real-axis spectra use Eqs.~(\ref{eq:nnls-real})-(\ref{eq:W-assembly})
evaluated on a grid $\omega\in[0,5.5]$\,eV with step $0.01$\,eV and $\eta=0.05$\,eV.

\paragraph*{Passivity and Kramers-Kronig checks.}
We enforce the passivity condition $-\mathrm{Im}\,\varepsilon^{-1}_{00}(\mathbf{q},\omega{+}i0^{+})\ge 0$
by clamping tiny negative values to zero. We then verify Kramers-Kronig consistency by
reconstructing the real part from the loss function,
\begin{equation}
  \mathrm{Re}\,\varepsilon^{-1}_{00}(\mathbf{q},\omega)
  = 1 - \frac{2}{\pi}\,\mathcal{P}\!\int_{0}^{\infty}
  \frac{\omega'\,\bigl[-\mathrm{Im}\,\varepsilon^{-1}_{00}(\mathbf{q},\omega')\bigr]}
       {\omega'^{2}-\omega^{2}}\,d\omega',
  \label{eq:KK}
\end{equation}
and reporting the relative RMS deviation against the model values \cite{Lucarini2005}.

\paragraph*{Dynamic versus baseline screening usage in rates.}
In all matrix elements we evaluate $W_{00}(\mathbf{q},\omega{+}i\eta)$ at bosonic
frequencies set by the electronic energy transfers,
\begin{equation}
  \hbar\omega_{41}=\varepsilon_{4}-\varepsilon_{1},\qquad
  \hbar\omega_{31}=\varepsilon_{3}-\varepsilon_{1},
  \label{eq:omega-legs-SI}
\end{equation}
with the $q$-partition~(\ref{eq:W-assembly}).

\begin{figure*}[ht]
\centering
\includegraphics[width=0.9\linewidth]{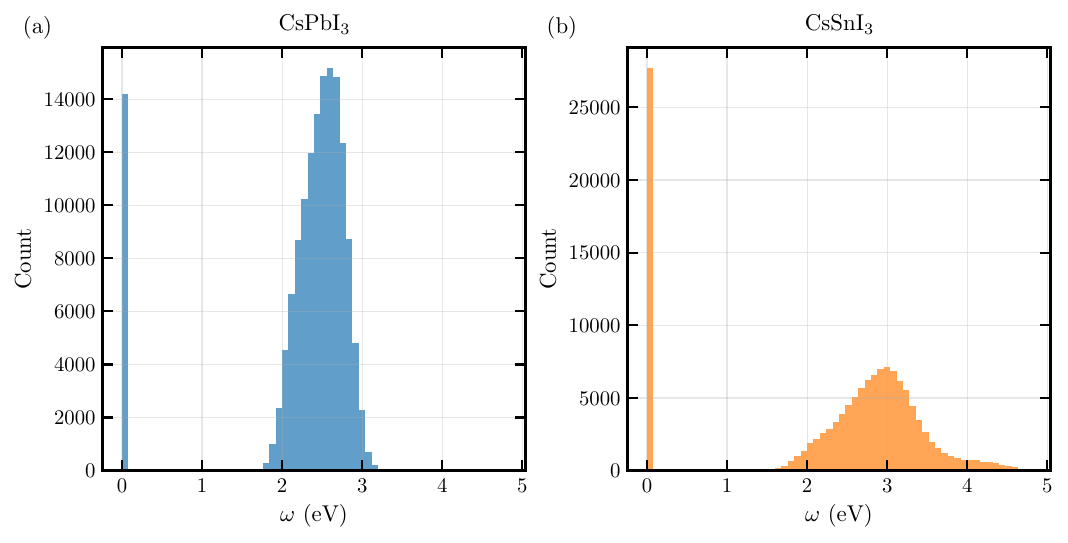}
\caption{Distribution of energy conserving quadruplets in $\omega$ sampled with the criterion $\Delta E \le 3\ \mathrm{meV}$ for (a) \ce{CsPbI3} and (b) \ce{CsSnI3}.}
\label{figS1}
\end{figure*}

\section*{S12. Frequency-space landscape of Auger-events}

To illustrate the distribution of energy- and momentum-conserving quadruplets, Fig.~S1 shows their spread in $\omega$. While there is a large peak near $|\omega|\!\approx\!0$, the distribution carries substantial finite-$\omega$ weight. Thus the clustering at $\omega\!=\!0$ is not representative of all sampled events, and we therefore evaluate $W_{00}(\mathbf q,\omega)$ at the event-specific energy transfers rather than adopt a kernel based on an $\varepsilon_\infty$ surrogate.

\section*{S13. Units and final outputs}

All rates $R$ are reported in $\mathrm{cm^{-3}\,s^{-1}}$ after multiplying the
cell-based expressions by $\mathrm{cm}^3/\Omega$. Coefficients $C_n$ and $C_p$
are in $\mathrm{cm^6\,s^{-1}}$ as in Eq.~(\ref{eq:CnCp}). Baseline/dynamic ratios,
material comparisons (CsSnI$_3$ \emph{vs.} CsPbI$_3$), $n$-$p$ grids, and
temperature trends are built directly from the $(T,n,p)$  outputs without interpolation.
All data used to construct the plots and draw inference are available at \cite{auger2025_data}.


\nocite{*}
\bibliography{main} 